# Temperature dependent Thermal conductivity of a single Germanium nanowire measured by Optothermal Raman Spectroscopy

Shaili Sett[1#], Vishal Kumar Aggarwal[1], Achintya Singha[2] and A. K. Raychaudhuri[1,3*]

[1]Department of Condensed Matter and Material Sciences, S. N. Bose National Centre for Basic Sciences, JD Block, Sector 3, Salt Lake, Kolkata 7000106, India.
[2]Department of Physics, Bose Institute, 93/1, Acharya Prafulla Chandra Road, Kolkata 700009, India
[3]Current address: CSIR-Central Glass and Ceramic Research Institute, S. C. Mullick Road, Kolkata 700032, India.
[#]Current address: Dept. of Physics, Indian Institute of Science, Kolkata 560012, India.
*Email: arupraychaudhuri4217@gmail.com

We investigate temperature dependent thermal conductivity $\kappa(T)$ in a single Ge nanowire (NW) using Optothermal Raman Spectroscopy which utilizes the temperature dependence of Raman lines as a local probe for temperature. The experiment was done from 300 K to above 700 K, a temperature range in which thermal conductivity of single NWs has been explored rarely. The thermal conductivity of Ge NWs (grown by vapor liquid solid mechanism), at around room temperature were observed to lie in the range 1.8 – 4.2 W/m.K for diameters between 50-110 nm. The thermal conductivity at a given temperature was found to follow a linear dependence on NW diameter, suggesting that the low magnitude of $\kappa(T)$ is determined by diffused scattering of phonons from the surface of NWs that reduces it severely from its bulk value. $\kappa(T)$ shows $\sim \frac{1}{T}$ behavior which arises from the Umklapp processes. The quantitative estimation of errors arising from the opto-thermal measurement and methods to mitigate them has been discussed. We also suggest a quick way to estimate approximately the thermal conductivity of Ge and Si NWs using the above observations.

## I. INTRODUCTION

In the last decade there has been intense interest on heat transport in nanostructures, in particular nanowires (NWs), for their applications in nanoelectronics and in specific areas like thermoelectric [1,2]. The experimental studies as well as theoretical investigations show that there can be large reduction of thermal conductivity from that of bulk in NWs when the size is reduced [3]. The reduction in thermal conductivity below a transverse length scale (like diameter in a NW) of 100 nm or below can arise from strong diffuse boundary scattering of phonons although this may not be the only cause. Effects of change in phonon dispersion relation and group velocity due to lateral confinement has also been proposed as a likely cause of reduction in thermal conductivity [4,5]. Recently, it has been experimentally demonstrated through Brillion—Mandelstam Spectroscopy that acoustic phonon group velocity decreases and there are changes in the phonon density of states in a nanowire of diameter as large as 128 nm [6].

A proper evaluation of thermal conductivity of nanostructures like NWs and nanolines is important from the view of two opposing requirements. In nanoelectronics and in optoelectronics using NWs, low thermal conductivity is a matter of concern as this may lead to thermal hot spots due to high current density, whereas for applications in thermoelectrics one would need to reduce the thermal conductivity.

Thermal conduction in Si NWs has been investigated extensively (for $T \leq 350$ K) using both experimental [7,8] and theoretical tools [9]. The main observation is that the thermal conductivity in Si NWs is low and can be even an order less than the bulk value [9]. The low thermal conductivity is believed to arise primarily from scattering of phonons at the boundary, which being diffusive in nature, reduces phonon lifetime [10].

The growth in research in experimental determination of thermal conductivity in NWs faces a bottleneck in measurements particularly above room temperature. At higher temperatures (for $T > 350$ K), due to heat loss by radiation, the measurements done are limited [11,12]. Determination of thermal conductivity ($\kappa$) of NWs, in particular on single NW, is nontrivial and often complex as these need use of micro fabrication techniques that can suspend NWs on $SiN_x$ or similar semiconductor membranes with microfabricated heater and thermometers [13]. This method has been used widely to determine the thermal conductivity of Si NWs, [7,8] Graphene [14], Carbon nanotubes [15] and other nanomaterials. With a custom designed thermally optimized sample stage, embedded radiation shield and better thermal anchor, it has been possible to measure thermal conductivity upto ~700 K [11,12]. Even though these methods are accurate, they need elaborate micro/nano fabrication. A rapid and simpler noncontact method, known as Optothermal Raman Spectroscopy was developed to measure $\kappa(T)$ initially for graphene and then other suspended low dimensional materials [16-18]. In this method, the Raman peak shift with temperature is used as a thermometer and the laser employed to collect the Raman Spectrum is used as a heat source. Application of this method in case of NW suffers from uncertainties that arise mainly from a proper estimation of power absorbed by the NW. We will discuss this in the relevant section.

Recently, the same method was used to measure the thermal conductivity of cantilevered Si NWs (clamped at one end) at 300 K [19]. The results lie within 5% -10% of thermal conductivity of Si NW measured using conventional electrical methods made on microfabricated platforms. The reported work, however, has not investigated the application of the method for temperatures above or below room temperature. This method is somewhat "accuracy challenged", [20] and requires the prior knowledge of absorption coefficient of the test material. As investigated in this work, we show that this method can also be used for $T$ well above 300 K and even for $T \geq 750$ K, which is a relevant temperature range for diverse applications.

The investigations on thermal transport in NWs have been mostly done on Si NWs [7,8,10]. In contrast thermal transport in single Ge NW with good structural quality has not been investigated extensively. In recent years Ge NWs have become a topic of considerable research in electronics [21] and optoelectronics [22]. There has been a report of an $Si_{0.14}Ge_{0.86}$ alloy NW of diameter 205 nm that shows a considerably low thermal conductivity of ~1.8 W/m.K at 300 K [23]. There is only one report on thermal conductivity of a single Ge NW from 100 K to 390 K through a microchip measurement setup. It quotes a value of 2.26 W/m.K for a 20 nm NW at 300 K [24] which is ~40% less than the predicted value of the same from theoretical consideration. This report of low thermal conductivity in Ge NW thus serves as one of the motivations to research it, so that its applicability in thermoelectrics can be tested. Evaluation of thermal conductivity in Ge NWs also become important in view of local hot spots that can originate from low thermal conductivity when such NWs are used in optoelectronic detectors where large current density ($10^8$-$10^9$ A/m$^2$) [22] may result from ultra-high responsivity observed in them giving rise to a local Joule heating.

In this work, thermal conductivity of a crystalline Ge NW was measured as a function of temperature from 300 K to above 700 K by Optothermal Raman spectroscopy. Results show the efficacy of size reduction (i.e. diameter reduction) in bringing about the reduction in thermal conductivity and also evaluates the temperature dependence of thermal conductivity and explores the applicability of such process as Umklapp process in NWs. We evaluated the results in the framework of physical theories of heat conduction in NWs. We have also estimated the uncertainty associated with this method.

## II. METHODS AND EXPERIMENTAL DETAILS

### II (i). Sample Preparation

Ge NWs were grown by physical vapor transport in a dual zone furnace using the vapor liquid solid mechanism and Au nanoparticle (NP) as seed catalyst on a Si substrate as described in our earlier publication [25]. NWs that grow near the edge of the substrate grow outward and are clamped at one end and are suspended in the other without any contact with the substrate as shown in Figure 1. These NWs have been used for measurements of thermal conductivity because, absence of any support at the free end avoids spurious heat transport path which contributes to a source of error in such thermal measurements.

Ge NWs used in the experiment are single crystalline and highly oriented as can be seen from the High Resolution Transmission Electron Microscope (HRTEM) image as well as Selective Area Electron

Diffraction (SAED) pattern (see Figure. 2(a) and 2(b)). SAED pattern shows its cubic structure. The $d$ plane spacing is ~0.331 nm which corresponds to the <111> plane. The surface of the Ge NW is covered by a native oxide layer (1-2 nm in thickness) as seen from the HRTEM image. The surface plays a crucial role in determining the dominant scattering route, especially in nanomaterials. The surface of a NW can be quantified through a roughness parameter $h$, which is the rms roughness at the surface. From atomic force microscopy, we have extracted the rms surface roughness of the NW which is ∼ 0.48 ± 0.05 nm.

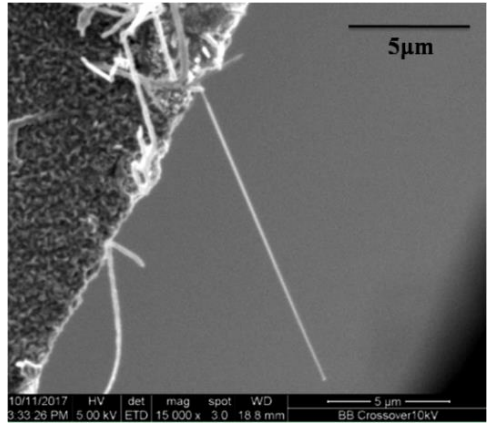

FIG. 1. SEM image of an edge of a Si substrate on which we have grown Ge NWs. The NW is clamped at one end and is suspended on the other, creating a cantilever.

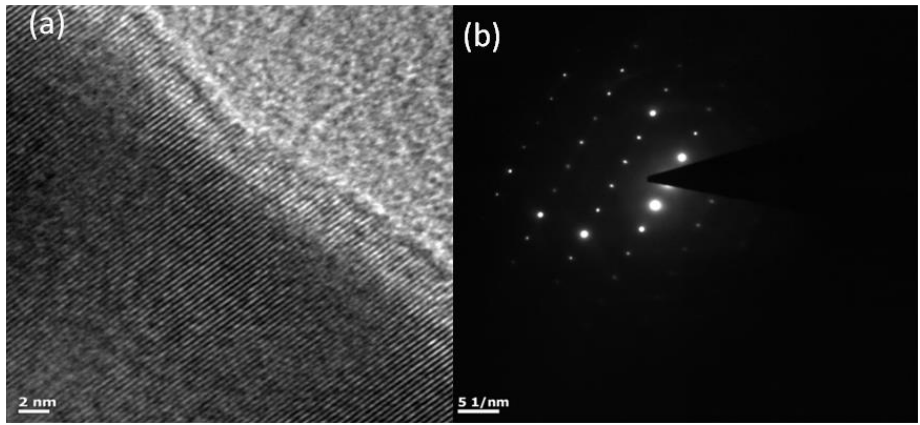

FIG. 2. (a) HRTEM image of a Ge NW showing highly oriented growth with parallel $d$ planes along <111> direction. (b) SAED pattern which establishes the crystalline nature of the NW.

## II (ii). Optothermal measurement

The heat flow in a cantilevered NW (clamped at one end and suspended in the other) when exposed to a heat flux $Q$ at a point at length $L$ from the clamped end can be expressed as a linear 1D equation, with the parameters, thermal conductivity ($\kappa$) and thermal contact resistance ($R_{th}$) given by [19],

$$\Delta T = QR_{th} + \frac{LQ}{A\kappa}, \tag{1}$$

where, the temperature drop $\Delta T \, (= T - T_o)$ is established over the length $L$ of the NW with cross sectional area $A$ and $T_o$ is the substrate (sink) temperature. The presence of thermal contact resistance $R_{th}$ at the NW/Si substrate interface creates a jump in temperature at the cold clamped end. The temperature $T^*$ at the cold end of the NW is given by,

$$T^* - T_0 = R_{th}Q \tag{2}$$

Determination of $R_{th}$ properly mitigates a major source of error. We have determined $R_{th}$ from experimental data using equation 1 and also using a finite element analysis [26, 27]. The heat flux is given as $Q = P\eta$, where $P$ is the power falling on the NW and $\eta$ is the absorption efficiency. The exact value of $\eta$ used in this work is discussed later.

The temperature gradient in the NW of length $L$ after correcting for the temperature rise at the contact is given by $\Delta T' = T - T^*$. The mean temperature at which thermal conductivity is measured is given by

$$T' = T^* + (\Delta T'/2). \tag{3}$$

In this method the temperature is measured from the shift of the Raman line which acts as a thermometer. First, we calibrate the Raman line shift with temperature. Ge NWs are dispersed on to a substrate which defines the temperature of the NW on which Raman shift will be measured. The temperature dependent Raman spectrum was recorded in a Lab RAM HR spectrometer with 1800 gr/mm grating and Peltier cooled CCD detector. The Raman data has been collected by focusing a 488 nm Argon ion laser through a 50X objective lens (FWHM of laser is ~1.4 μm) at a low laser power that gives a good signal to noise ratio yet avoids any local heating that may give rise to peak broadening and/or peak shift. Typical laser power used was $\leq 2$ μW and for the NW with smallest diameter power used is ~ 1 μW. The Raman spectra peak is fit using a Lorentzian function to extract the peak position (see Figure 3). The peak position at room temperature ~ 300.5 cm$^{-1}$ is from the degenerate LO/TO mode at $q = 0$ [28]. A bulk Ge Raman spectrum fitted to a Lorentzian has also been shown in Fig. 3. The diameter of the NWs is $\geq 50$ nm and there are no effects of phonon confinement as observed from the spectra. Phonon confinement is observed in thinner NWs with diameter < 20 nm [29], where asymmetries in the peak shape is observed. We have also measured the Raman spectrum of a single suspended Ge NW at room temperature (see

Figure 3). The results are identical, and the temperature equilibrated thermopower does not depend on whether the NW is on a substrate or is suspended.

The temperature variation of the substrate from 250 K to 500 K was carried out in a vacuum chamber connected to a PID controlled heating stage, with $N_2$ vapor cooling facility. We have recorded the Stokes line in this experiment. Drift in the stability of the temperature controlled stage leads to defocusing which can be detected from drift of the Raman line. The data was taken only after the temperature of the stage shows good stability. An example of the Raman line at different temperatures is given in Figure 4(a). The calibration of the Raman peak shift with temperature was performed on a single Ge NW which was dispersed onto a Cu grid that has been anchored to the temperature controlled stage. We get the peak position as a function of temperature from this measurement. This calibration acts as a temperature sensor for the NW as shown in Figure 4(b). We can observe the peak shift to a lower wavenumber as temperature of the NW increases. The peak position ($\omega$) as a function of temperature ($T$) is shown in Figure 4(b) for a single Ge NW. The slope of the straight line, $\Delta\omega/\Delta T = 0.0111\pm0.0004$ cm$^{-1}$/K serves as a calibrated thermometer. It is used to determine the temperature at any point in the NW through the peak position.

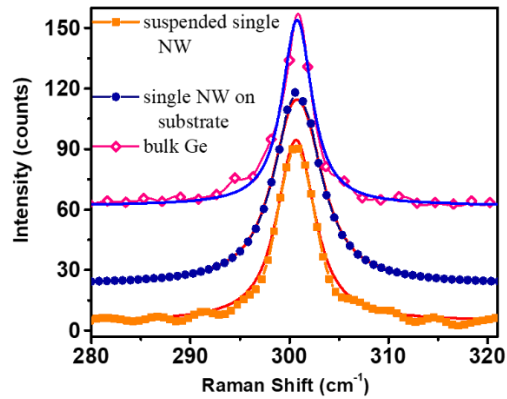

FIG. 3. Raman Spectrum of bulk Ge, single Ge NW suspended and supported on Si substrate with a peak at ~300.5 cm$^{-1}$ fit to a Lorentzian function.

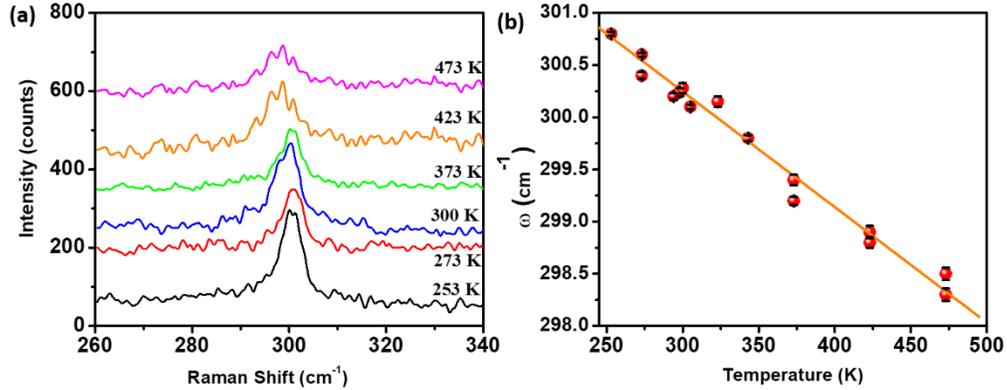

FIG. 4. (a) Some representative Raman spectra as a function of temperature in a single Ge NW. (b) Peak position versus temperature in a single Ge NW, with solid line showing the linear fit to data.

A realization of the utilization of equation 1 for measurement of the thermal conductivity can be done by measuring the Raman spectrum at a fixed power ($Q$) as a function of length of the distance of the point of illumination from the base of the NW ($L$). The slope of the $\Delta T$ v/s $L$ is then used to evaluate thermal conductivity. $\kappa(T)$ in this case is evaluated at the average temperature $T'$ (see Equation.3). To avoid errors, $\Delta T$ was kept small. This method does not need determination of $R_{th}$ but allows us to determine it through equation (1).

In another realization of measurement using eqn. (1) we varied the heat flux $Q$ at a fixed length $L$. $\kappa(T)$ was determined from slope of the observed $\Delta T$ v/s $Q$ curve. Although this method reduces the uncertainty in determination of $L$, it needs knowledge of $R_{th}$.

One of the important parameters for our experiments is the value of absorption efficiency ($\eta$) in the nanowire. The absorption efficiency of NWs is defined as the ratio between the absorption cross section and the projected area of the NW. It uses Mie solutions of the Maxwell equations, [30] allowing us to determine the energy absorbed by a NW. It depends upon the NW diameter and laser wavelength only. Absorption efficiency in a single nanowire is a source of uncertainty mainly because it is difficult to determine it in a freely suspended/cantilevered geometry. Measurements in nanowires have been done in ensembles or arrays supported by a substrate. This can also change the value of $\eta$ [31]. The value of absorption efficiency in a single NW thus has been mostly determined by simulation [32-34]. In Ge NWs in the diameter range that we are working the contributions made by resonant modes in the NW occur for wavelength > 600 nm [33,34]. For wavelength < 600 nm (that encompasses the range of 50-110 nm that we are working in) the absorption and scattering are independent of the diameter of the wire. The spectral dependence of absorption efficiency has been done by L. Cao et.al, [33] by direct measurement of the photocurrent in a single NW over a broad spectral range down to 500 nm. We have also measured the spectral response of a single Ge NW [35, 36] over the spectral range down to 300 nm. There is a broad agreement of the current measured by us with Cao and coworkers [33]. These measurements along with simulation gives an estimate of $\eta \approx 0.8 \pm 0.1$ for wavelength $\lambda = 500$ nm. From the spectral dependence of photocurrent in the single Ge NW measured we find that in the range from 400 nm to 500 nm the photocurrent changes by only 2% for a change in wavelength of 10 nm. This being within the uncertainty of the value of $\eta$, we thus assume a value of $\eta \approx 0.8 \pm 0.1$ for our measurement done at $\lambda = 488$ nm.

Since in the wavelength range ($\lambda < 600$ nm) the resonance modes donot contribute to optical response we did a simulation [37] to find the efficiency of absorption in the same geometry as ours. We find $\eta \approx 0.75$ which is within the range of value $\eta \approx 0.8 \pm 0.1$ found before.

For the determination of $\kappa(T)$, a single suspended NW is identified through a 100X microscopic objective and an Argon ion laser of 488 nm is focused on the NW through the lens with a 0.9 numerical aperture. An optical microscope image of the suspended NW with a focused laser is shown in Figure 5. The laser beam falling on the NW has the form of a Gaussian given by, $J = J_0 \frac{e^{\frac{(x-xo)^2}{2\sigma^2}}}{\sqrt{2\pi}\sigma}$, where $x$ direction is along the length of the NW, $xo$ is the point at which the laser beam falls (center of beam), $\sigma$ is the half width at half maxima. The power $P$ (total heat falling on the NW) is calculated from the intensity of the laser beam $I$, given by, $I = J/\pi\sigma^2$, where $J$ measured through the flux meter. Thus, the power on the NW can be approximated as, $P = I.S$, where $S$ is the area of the NW exposed to illumination. The beam falls upon $2\sigma$ length of the NW (along x axis). While in the perpendicular direction to the axis of the NW, the beam falls only on $\frac{\pi D}{2}$ region. So, $S = \frac{\pi D}{2}.2\sigma$ giving, $P = \frac{J}{\pi\sigma^2}\frac{\pi D}{2}.2\sigma$

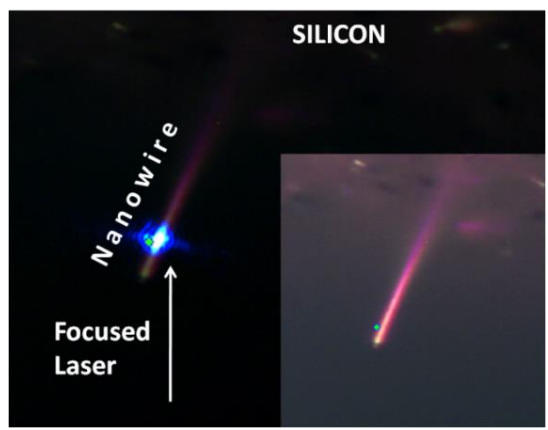

FIG. 5. A 488 nm Ar ion laser focused on a single cantilevered Ge NW. Inset showing the suspended NW without laser illumination.

Focusing the light on the nanowire also constitutes a source of error. The beam area ($\pi\sigma^2$ ~0.38μm²) is larger than the area of the NW exposed to illumination ($\frac{\pi D}{2} \cdot 2\sigma$ ~0.11 μm²). The laser light was focused for each measurement to preserve the Gaussian profile at the excitation wavelength. The beam as shown in Figure 5 looks like an ellipsoid as the distance of illumination along length is larger than the illumination along the diameter. For each measurement, the laser was focused multiple times at a particular position on the NW till the peak shift of the Raman line was reproducible and also of lowest value as it signifies largest temperature gradient created at a given power at a point of illumination. Errors arising from improper focusing have been minimized by monitoring the parameters like beam shape falling on the NW, reproducibility of data on refocusing and absence of large noise.

One of the sources of error in applying Equation (1) arises from finite size of the of heat source ((i.e., the finite size of the laser spot). It is not a point source but can be approximated as one when the width of the illumination region ($2\sigma$) $\ll L$. Using this information and the fact that > 95% of the power is contained within an area of diameter $2\sigma$, the effective size of the illumination region is ~ 0.7 μm. For measurement of thermal conductivity, we used a length of wire $L > 15$ μm, which makes the error arising from this source $\leq 10\%$. Validation of point source approximation is shown through finite element method simulations [38].

Another source of error in high temperature thermal conductivity experiments is heat loss through convection and radiation as discussed in Section 1. We have used finite element method to determine the heat loss through radiation processes [38]. The radiation heat loss in a Ge NW at temperatures ~ 600 K is ~ 3 nW which is ≤ 2 % of the total heat carried by the NW. However, the heat loss through convection and conduction can be appreciable, in particular when experiments are done at higher temperatures which need use of higher power [39]. Briefly, the following observations have been made. The difference between the data taken in air and in vacuum are not appreciable for lower powers (difference in gradient < 10%, for power ≤ 5 μW and < 15% for power ≤ 7.5 μW.) This is the power used for measurements up to 400 K. Thus, data of thermal conductivity at or close to room temperature, do not get affected significantly when data are taken in air as compared to that taken in vacuum. However, at higher power (> 8 μW) the difference in these measurements can be appreciable. At the highest power used by us (14 μW), the difference in gradients for measurements in vacuum and in air ~ 30%. The data presented here are corrected for these effects as stated above.

### III. RESULTS

**III (i). Estimation of thermal conductivity through length variation method**

An example of a Raman spectrum (with local heating) in NW B for varying positions of the laser spot on the NW is shown in Figure 6(a). In the length variation method, since the active $L$ varies from 2 – 20 µm, we use very low power ($J < 2$ µW) to avoid heating up regions lying outside the $2\sigma$ region. For such a low power the difference in data the taken in air and in vacuum are negligible [38, 39]. This ensures that we maintain the condition that $L \gg 2\sigma$ and minimize errors arising from $L$ calculation. The active length $L$ which is the distance of the illumination spot from the point of clamp on the substrate was obtained from $x\,y$ position markers accompanying the movable sample stage. The uncertainty in length comes from drift due to the sample stage vibrations. The temperature gradient $\Delta T$ is obtained from our calibrated Raman thermometer using $\Delta \omega / \Delta T = 0.0111 \pm 0.0004$ cm$^{-1}$/K and the plot of $\Delta T$ versus $L$ (Figure 6(b)) is obtained. Using the slope of this line, we can calculate the thermal conductivity of the NW using Equation 1. The determination of the $R_{th}$ is not needed in the length variation method and we calculate $\kappa$ from the slope, although $R_{th}$ can be evaluated from the method (see Table I). The estimated $R_{th}$ is similar for all the contacts ~ 0.4 - 0.5 K/nW. This number gives us an estimate of temperature rise ($T^* - T_0$) at the contact which is ~ 90 K to 25 K for the $P$ used. (Note: The Raman peaks in Figure 6 are fitted with a Lorentzian to find their exact peak positions. The small asymmetry seen arises from local heating [29]. The fitted curve to a Lorentzian function [40].

Table I gives the essential fit parameters to Equation 1. The temperature $T'$ at which the thermal conductivity is measured has been calculated using Eqn. (3). Within the total $\Delta T$ of ~100 K, $\kappa$ varies by $\leq 10\%$. Thus, we assume that $\kappa$ is independent of temperature within this temperature range. From the data in Table I we can get a quantitative estimate of variation of $\kappa$ with diameter. (Note: Approximate relative variation in $\kappa$ within the range 405 K – 440 K is < 5 % change. This is less than the uncertainty in the $\kappa$ data.) The thermal conductivity data in this range is taken with lower power as stated earlier so that the data taken in air and in vacuum do not differ much and are within the measurement uncertainties as shown in Table I.

The thermal conductivity is plotted in Figure 7 as $\kappa$ versus $D$ and linear dependence suggests that the thermal conductivity is dominated by boundary scattering. The boundary scattering time ($\tau_B$) term is defined as, [41]

$$\tau_{B,j}^{-1} = \frac{v_j}{D}\left(\frac{1}{F}\right), \tag{4}$$

where $v_j$ is the cumulative group velocity of the $j$<sup>th</sup> mode of a phonon, $D$ is the diameter and $F$ is the Rugosity Factor [41] which is the ratio of diffuse to specular reflection. This observation is important because it allows us to estimate $\kappa$ at any diameter from measurement made on NWs as discussed later.

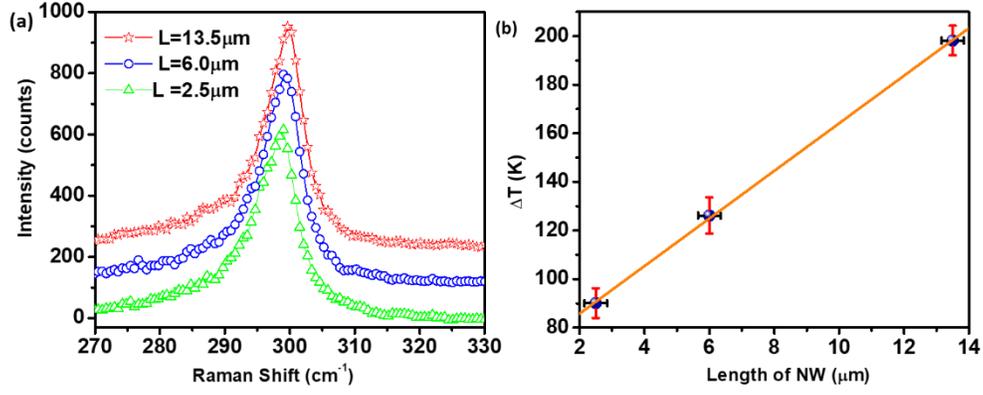

FIG. 6. (a) Magnified Raman spectrum as a function of length (*L*) for a fixed power in NW B. (b) *ΔT* as a function of length for NW with linear fit to data in solid line. The incident power for this measurement is < 2 μW.

TABLE I. Ge NW with fit parameters to Equation 1 for the length variation method.

| NW | *D* (nm) | *J* (μW) | *P* (nW) | $R_{th}$ (K/nW) | *T′*(K) | *κ* (W/m.K) |
|---|---|---|---|---|---|---|
| A | 110 | 1.85 | 185±5 | 0.42±0.01 | 426 | 3.8±0.5 |
| B | 72 | 1.93 | 126±5 | 0.52±0.02 | 440 | 2.6±0.5 |
| C | 50 | 1.04 | 48±5 | 0.46±0.05 | 405 | 1.9±0.4 |

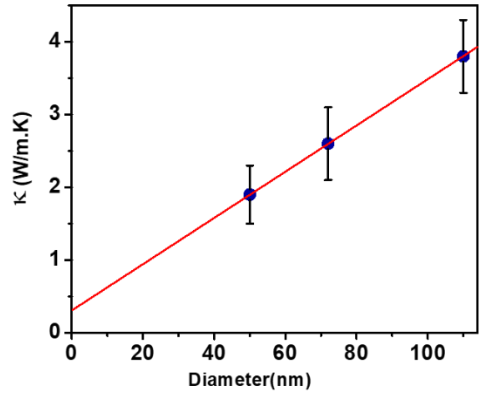

FIG. 7. Thermal conductivity of a single Ge NW as a function of diameter.

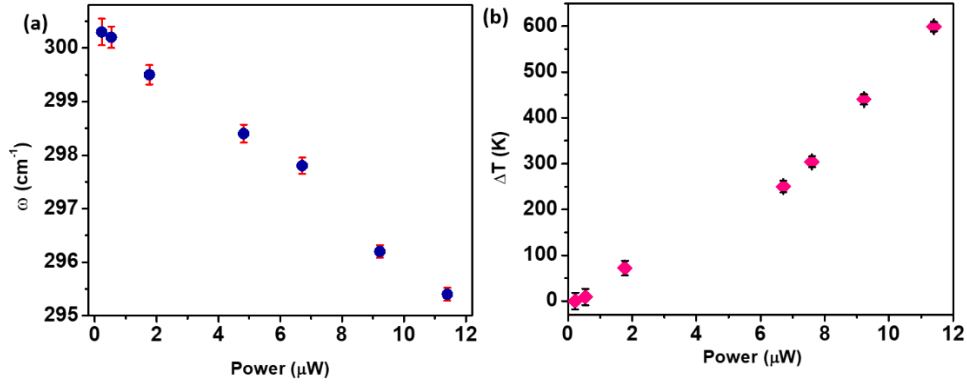

FIG. 8. (a) Peak positions as a function of laser power for a fixed length in NW A. (b) $\Delta T$ as a function of power for NW A.

### III (ii). Estimation of thermal conductivity through power variation method

The Raman spectra for a fixed length $L$ from the laser beam to the base of the NW for varying powers were recorded for two NWs, A and B. The peak position as a function of laser power is plotted in Figure 8(a). As we increase power, the local temperature at position $L$ increases leading to a peak shift $\Delta \omega$. From the $\Delta \omega$ we calculate $\Delta T$ as a function of laser power as shown in Figure 8(b). We use simulation results [27, 38] as well as the experimental result (see Table I) to calculate the corrected $\Delta T'$ using the value of $T$. We use Equation 1 to determine the thermal conductivities of the NWs as function of temperature $T'$ shown in Figure 9. The results have been corrected for high powers due to effects of heat transfer through air surrounding the NW [39]. The data shows an inverse dependence of $\kappa$ on $T$ discussed in detail in the next subsection.

The magnitude of $\kappa$ of the Ge NWs ~ 4 W/m.K (diameter 110nm) is less than that of the single crystalline Ge by about a factor of 15 at room temperature and for the 50 nm diameter NW is lower by a factor of 30. This is an important quantitative evaluation. In addition, as pointed out before, we also make two important observations that $\kappa$ has an inverse dependence on $T$ and a linear dependence on $D$. In the next section we discuss the physical significance of these results.

### IV. DISCUSSIONS

#### IV(i). Temperature dependence of thermal conductivity

In the temperature range $T > \theta_D$ (Debye temperature), $\kappa(T)$ of an insulating crystalline solid generally shows $1/T$ dependence which arises from inelastic Umklapp processes [42]. The inverse $T$ dependence of $\kappa$ observed in these NWs thus can be postulated to arise from Umklapp process. To extract the temperature dependence in $\kappa(T)$, we have fitted the thermal conductivity of NWs to an equation of the form,

$$\kappa^{-1}(T) = {\kappa_o}^{-1} + {\kappa_U}^{-1}(T), \qquad (5)$$

where $\kappa_U(T) = \frac{b}{T}$ is the temperature dependent term due to Umklapp scattering. Generally, phonons have a wide distribution of mean free paths and to substitute it with Equation 5 may be incorrect. But in this case, we have followed several simulations and found that the dominant contribution at room temperature and above comes from long wavelength phonons. Unless we have a 1D type Density of states, (which occurs when the diameter of the NW falls below 10nm), these long wavelength phonons dominate. So, approximation that at higher temperature Umklapp process (three phonon scattering) can be characterized by phonon of a dominant wavelength is acceptable.

The temperature independent part $\kappa_o$ is a constant arising from elastic scattering of phonons such as boundary scattering. The transformed equation 5 in terms of $\kappa(T)$ is,

$$\kappa(T) = \frac{\kappa_o b/T}{\kappa_o + b/T} \qquad (6)$$

The fit parameters to Equation 6 are given in Table II and the fit data is shown in Figure 9. In the following section we evaluate the postulates quantitatively to establish its correctness.

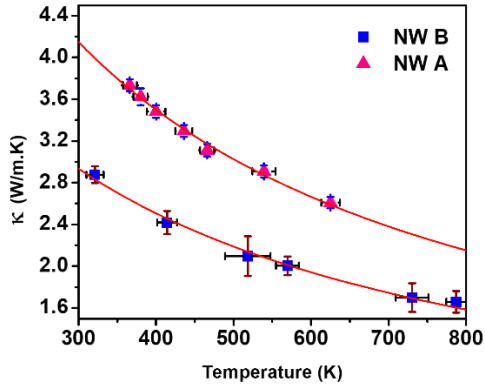

FIG. 9. Thermal conductivity as a function of temperature in the NWs A and B. Solid lines are fit to Equation 6.

If the inverse temperature dependence indeed arises from the Umklapp process, (see Equations 5 and 6) then, $\kappa_U$ for all temperatures above the Debye temperature can be expressed as [42],

$$\kappa_U = \frac{\rho a v_c^3}{\gamma^2 T}, \ (T > \theta_D) \qquad (7)$$

where $\rho$ is the density, $a$ is the lattice constant, $\gamma$ is the Gruniesen parameter and $v_c$ is the cumulative sound velocity in the material. The observed values of $b$ can be used for finding $v_c$ for the NWs. We use the standard macroscopic values of Ge i.e., $\rho$ = 5.32 g/cm$^3$ and $\gamma$ = 1.4. The lattice constant of Ge NW can be evaluated from the $d$ plane spacing (from TEM study) [25] of 0.331 nm, which gives $a$ = 5.73Å. We find that for the Ge NW with diameter of 110 nm, $v_c \sim$ 1129 m/s and for the Ge NW with diameter of 72 nm, $v_c \sim$ 1035 m/s, while in bulk Ge, $v_c \sim$ 2575 m/s [41]. (Note: Determination of the cumulative velocity $v_c$ of bulk Ge have been done using the equation: $\frac{1}{v_c^5} = \frac{1}{v_l^5} + \frac{2}{v_t^5}$, where $v_l$ is the longitudinal mode velocity and $v_t$ is transverse mode velocity [43].)

Considering that there is no adjustable parameter in Equation 7 we think that the value of $v_c$ as obtained from the fit, is a reasonably good agreement with other published data. It has been observed that the sound velocity softens for NWs, even for diameters ~ 120 nm [6]. There are reports of observations of softening of Debye temperature (a quantity directly related to sound velocity) even in metallic nanowires [44].

Based on the above discussion and the quantitative reasonableness of the parameters obtained from Umklapp theory it can be established that the inverse temperature dependence of the thermal conductivity ($\kappa \sim \frac{1}{T}$) indeed arises from the Umklapp process.

TABLE II. Fit parameter to Equation 6 shown in the $\kappa$ versus $T$ plot.

| NW | Diameter ($D$) (nm) | $\kappa_o$ (W/m.K) | $b$ (W/m) | $v_c$ (K) |
|---|---|---|---|---|
| A | 110 | 9.4±0.7 | 2236±98 | 1129±9 |
| B | 72 | 6.0±0.4 | 1723±80 | 1035±8 |

### IV (ii). Boundary scattering and thermal conductivity

The temperature dependent part as discussed in the sub section before arises from the Umklapp process that is comparable to what is observed in crystalline bulk albeit with reduced sound velocity. However, the large part of thermal conductivity reduction in NWs appears to arise from scattering (diffused) at the NW boundary and is an elastic process. Previous investigations on ultrathin Si films /Si NWs and associated Monte Carlo [45]/MD simulations [46] have shown that the major cause for reduction in thermal conductivity in such size reduced systems can indeed be diffused scattering from the boundary.

To check the contribution of the boundary scattering quantitatively, we have calculated the thermal conductivity of Ge NWs using the modified Callaway formalism. The thermal conductivity is given by, [8,10,47]

$$\kappa(T) = \sum_j \frac{k_B}{8\pi^3 v_j} \left(\frac{k_B T}{\hbar}\right)^3 (I_j) \tag{8}$$

where the summation index $j$ is over all modes, $v_j$ is the group velocity of the $j^{th}$ mode and other symbols have their usual meaning. The total thermal conductivity due to a longitudinal ($j = L$) mode and two transverse modes ($j = T$) gives,

$$\kappa = \kappa_L + 2\kappa_T. \tag{9}$$

The integral $I_j$ are defined as,

$$I_j = \int_0^{\theta_{D,j}/T} \int_0^{2\pi} \int_0^{\pi} \tau_{B,j} \frac{x^4 e^x}{(e^x-1)^2} Cos^2\theta Sin\theta d\varphi dx \tag{10}$$

where $\theta_{D,j}$ is the Debye temperature for a particular mode $j$. The boundary scattering time ($\tau_{B,j}$) has been defined previously in Equation 4. In this approximation for completely diffusive scattering ($F=0$) and $\tau_B^{-1} \to \infty$. Apart from $F$ which is obtained from the fit to the data, the other parameters like $\theta_D$ has been taken from bulk value and $v_j$ is the sound velocity of the $j^{th}$ mode obtained experimentally.

TABLE III. Comparison of thermal conductivity from boundary scattering.

| Diameter $(D)$ (nm) | $\kappa$ (W/m.K) at 300K from fit to eqn. 10 | $\kappa_o$ (W/m.K) from fit to eqn. 5 |
|---|---|---|
| 110 | 9.39 for $F = 0.228$ | 9.4±0.7 |
| 72 | 5.94 for $F = 0.191$ | 6.0±0.4 |

From Table III it can be seen that there is an excellent agreement with the observed value (within the experimental uncertainty) which is obtained for $F \sim 0.2 \pm 0.03$. The finite but low value of $F$ can arise from acoustic impedance mismatch between Ge and GeO$_2$ leading to scattering at interface of Ge and surface GeO$_2$. More importantly we found from microscopy data that the surface has a corrugation length scale ~10 nm and height corrugation ±2 unit cells. As a result, finite and low $F$ can arise. From previous molecular dynamics simulation of $\kappa$ done on Si NWs with rectangular cross section with area of cross section varying from 2.58 nm$^2$ to 28.6 nm$^2$ and its comparison with calculations based on Boltzmann

Transport Eqn. with boundary scattering, it was found that the two agree when the specularity factor $F = 0.45$ [46]. This difference can arise because Ge NW used by us has a rather larger area of cross section. This increases the amount of specular reflections from the boundary as compared to the diffuse scattering and reduces $F$. Also, the materials are different, and the surface conditions are different.

## IV (iii). Predictability of thermal conductivity for NWs

Predicting thermal conductivities of NWs without measurements will be a useful proposition even if it is done with some degree of known uncertainties because its measurement is quite complex, and it is not possible to carry it out whenever some estimate of $\kappa$ is needed. The results presented and the subsequent analysis allows us to predict thermal conductivity of semiconductor NWs with good crystallinity albeit with some degree of uncertainties. Given the fact that the thermal conductivity in these NWs is dominated by phonon conductivities, (due to its low electrical conductivity) boundary scattering and Umklapp scattering can be taken as the main source of phonon scattering for $T > \theta_D$ which is a valid approximation for most NWs at $T \sim 300$ K. Presence of electrons (in case of doped NWs) will provide a source of extra phonon scattering due to presence of dopants and also due to phonon electron scattering which will likely decrease its thermal conductivity from the intrinsic value. Presence of lattice defects in case of these NWs that has less structural quality will have similar effects. The boundary scattering contribution can be estimated from a specularity factor of $F \approx 0.2$ which is a reasonable estimate for NWs. We do have a caveat that a much larger value of $F$ as may arise in NWs that have very rough surfaces (like those made by plasma or chemical etching). We find that the temperature dependent part is weak and can be obtained from Equation 7 of Umklapp scattering contribution with uncertainty though not very high from reduction of sound velocity from the bulk value in the NW.

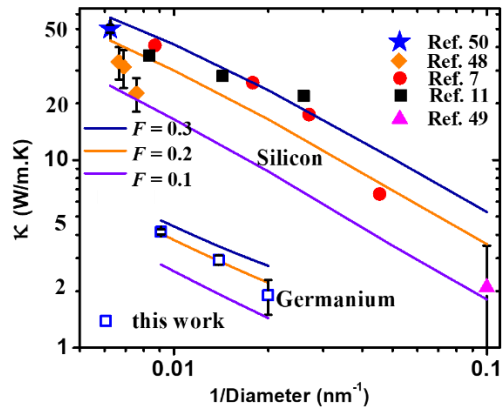

FIG. 10. Predictability of thermal conductivity of NWs at 300K with experimental data points of Si NWs from Ref [7, 11, 48-50].

We may thus suggest that Equation 6 can be used to get a prediction of thermal conductivity of individual intrinsic (or lightly doped) semiconductor NWs with a given diameter where the term $\kappa_0$ can be estimated from boundary scattering (Equation 8) and the term $\frac{b}{T}$ estimated from the Umklapp process (Equation 7). We have used these to estimate the value of $\kappa(T = 300\ K)$ for $0.1 \leq F \leq 0.3$ in Si and Ge NWs and the results are shown in Figure 10. Alongside we have also shown experimental data in Si NWs [7,11,48-50] and Ge NWs from our work. The estimated uncertainty is $\pm\ 20\%$. This predictability when validated on different NWs may provide a tool to estimate this quantity in narrow semiconductor NWs without measurements. It is noted that this estimate has been done using simple classical thermal conductivity theories, but it does provide a quick estimate although with not so small an uncertainty ($\sim \pm\ 20\%$).

## V. CONCLUSIONS

In summary, we have measured thermal conductivity in single cantilevered Ge NWs through Optothermal Raman spectroscopy utilizing the Raman line shift with temperature as a calibrated temperature sensor. The thermal conductivity of the NWs is almost an order less than Si NWs of the same dimension. The thermal conductivity of the NWs shows a linear dependence on diameter as well an inverse dependence on temperature, both of which have been quantitatively discussed. The thermal conductivity above Debye temperature is governed by two important scattering mechanisms: boundary scattering and Umklapp scattering. Armed with these two quantities that can be approximated from bulk values (making suitable modifications in parameters as are expected in NWs), the thermal conductivity of semiconductor NWs with different diameters can be predicted within $\pm$ 20% uncertainty.

**ACKNOWLEDGEMENTS**

AKR acknowledges financial support from Science and Engineering Research Board, Government of India as J.C. Bose Fellowship (SR/S2/JCB-17/2006) and SERB Distinguished Fellow (SB/DF/008/2019). The authors acknowledge the support of Chandan Samanta and Prof. Barnali Ghosh, S. N. Bose National Centre for Basic Sciences, Kolkata.

**REFERENCES**

[1] D. G. Cahill, W. K. Ford, K. E. Goodson, G. D. Mahan, A. Majumdar, H. J. Maris, R. Merlin and S. R. Phillpot, Nanoscale thermal transport, J. Appl. Phys., 93, 793 (2003).

[2] D. G. Cahill, P. V. Braun, G. Chen, D. R. Clarke, S. Fan, K. E. Goodson, P. Keblinski, W. P. King, G. D. Mahan, A. Majumdar, H. J. Maris, S. R. Phillpot, E. Pop, and L. Shi, Nanoscale thermal transport II., Appl. Phys. Rev., 1, 011305 (2014).

[3] W. Fon, K. C. Schwab, J. M. Worlock and M. L. Roukes, Phonon scattering mechanisms in suspended nanostructures from 4 to 40 K, Phys. Rev. B, 66, 045302 (2002).

[4] A. Balandin and K. L. Wang, Significant decrease of the lattice thermal conductivity due to phonon confinement in a free-standing semiconductor quantum well, Phys. Rev. B, 58, 1544-1549 (1998).

[5] J. Zou and A. Balandin, Phonon heat conduction in a semiconductor nanowire, J. Appl. Phys., 89, 2932-2938 (2001).

[6] F. Kargar, B. Debnath, J. Kakko, A. Säynätjoki, H. Lipsanen, D. L. Nika, R. K. Lake and A. A. Balandin, Direct observation of confined acoustic phonon polarization branches in free-standing semiconductor nanowires, Nature Comm., 7, 13400 (2016).

[7] D. Li, Y. Wu, P. Kim, Li Shi, P. Yang and A. Majumdar, Thermal conductivity of individual silicon nanowires, Appl. Phys. Lett., 83, 2934 (2003).

[8] R. Chen, A. I. Hochbaum, P. Murphy, J. Moore, P. Yang, and A. Majumdar, Thermal Conductance of Thin Silicon Nanowires, Phys. Rev. Lett., 101, 105501 (2008).

[9] M. Kazan, G. Guisbiers, S. Pereira, M. R. Correia, P. Masri, A. Bruyant, S. Volz, and P. Royer, Thermal conductivity of silicon bulk and nanowires: Effects of isotopic composition, phonon confinement, and surface roughness, J. Appl Phys., 107, 083503 (2010).

[10] J. Lim, K. Hippalgaonkar, S. C. Andrews, A. Majumdar and P. Yang, Quantifying Surface Roughness Effects on Phonon Transport in Silicon Nanowires, Nano Lett., 12, 2475-2482 (2012).

[11] J. Lee, W. Lee, J. Lim, Y. Yu, Q. K. Jeffrey, J. Urban and P. Yang, Thermal Transport in Silicon Nanowires at High Temperature up to 700 K, Nano Lett., 16, 4133- 4140 (2016).

[12] X. Wang, J.n Yang, Y. Xiong, B. Huang, T. T Xu, D. Li and D. Xu, Measuring nanowire thermal conductivity at high temperatures, Measurement Science and Technol., 29, 025001 (2018).

[13] Li Shi, D. Li, C. Yu, W. Jang, D. Kim, Z.Yao, P. Kim and A. Majumdar, Measuring Thermal and Thermoelectric Properties of One-Dimensional Nanostructures Using a Microfabricated Device, J. Heat Transfer., 125, 881-888 (2003).

[14] W. Jang, W. Bao, L. Jing, C. N. Lau and C. Dames, Thermal conductivity of suspended few-layer graphene by a modified T-bridge method, Appl. Phys. Lett., 103, 133102 (2013).

[15] L. Lu, W. Yi, and D. L. Zhang, 3ω method for specific heat and thermal conductivity measurements, Rev. Sci. Instrum., 72, 2996 (2001).

[16] A. A. Balandin, S. Ghosh, W. Bao, I. Calizo, D. Teweldebrhan, F. Miao and C. N. Lau, Superior Thermal Conductivity of Single-Layer Graphene, Nano Lett., 8, 902-907 (2008).

[17] A. A. Balandin, Thermal properties of graphene and nanostructured carbon materials, Nat. Mater., 10, 569-581 (2011).

[18] H. Malekpour and A. A. Balandin, Raman-based technique for measuring thermal conductivity of graphene and related materials, J. Raman Spectroscopy, 49, 106-120 (2018).

[19] G. S. Doerk, C. Carraro, and R. Maboudian, Single Nanowire Thermal Conductivity Measurements by Raman Thermography, ACS Nano, 4, 4908-4914 (2010).

[20] T. Beechem, L. Yates and S. Graham, Invited Review Article: Error and uncertainty in Raman thermal conductivity measurements, Review of Scientific Instruments, 86, 041101 (2015).

[21] D. Wang, Q. Wang, A. Javey, R. Tu, H. Dai, H. Kim, P. C. McIntyre, T. Krishnamohan, and K. C. Saraswat, Germanium nanowire field-effect transistors with $SiO_2$ and high-k $HfO_2$ gate dielectrics, Appl. Phys. Lett., 83, 2432-2434 (2003).

[22] S. Sett, A. Ghatak, D. Sharma, G. V. Pavan Kumar, and A. K. Raychaudhuri, Broad Band Single Germanium Nanowire Photodetectors with Surface Oxide-Controlled High Optical Gain, J. Phys. Chem. C., 122, 8564-8572 (2018).

[23] Z. Wang and N. Mingo, Diameter dependence of SiGe nanowire thermal conductivity, Appl. Phys. Lett., 97, 101903 (2010).

[24] M. C. Wingert, Z. C. Y. Chen, E. Dechaumphai, J. Moon, J. H. Kim, J. Xiang and R. Chen, Thermal Conductivity of Ge and Ge-Si Core-Shell Nanowires in the Phonon Confinement Regime, Nano Lett., 11, 5507-5513 (2011).

[25] S. Sett, K. Das, and A. K. Raychaudhuri, Investigation of factors affecting electrical contacts on single germanium nanowires, J. Appl. Phys., 121, 124503 (2017).

[26] K. R. Hahn, M. Puligheddu and L. Colombo, Thermal boundary resistance at Si/Ge interfaces determined by approach-to-equilibrium molecular dynamics simulations, Phys. Rev. B, 91, 195313 (2015)

[27] See Supplemental Material at [URL will be inserted by publisher] for finite element analysis to determine thermal contact resistance at the Si/NW interface.

[28] J. H. Parker Jr., D. W Feldman and M Ashkin, Raman Scattering by Silicon and Germanium, Phys. Rev., 155, 712 (1967).

[29] R. Jalilian, G. U. Sumanasekera, H. Chandrasekharan and M. K. Sunkara, Phonon confinement and laser heating effects in Germanium nanowires, Phys. Rev. B, 74, 155421 (2006).

[30] M. Kerker, The Scattering of Light and Other Electromagnetic Radiation, 1st Edition, Physical Chemistry: A Series of Monographs, 13 (1969).

[31] R. Frederiksen, G. Tutuncuoglu, F. Matteini, K. L. Martinez, A. Fontcuberta-i-Morral and E. Alarcon-Llado, Visual Understanding of Light Absorption and Waveguiding in Standing Nanowires with 3D Fluorescence Confocal Microscopy, ACS Photonics, 4, 2235-2241 (2017).

[32] R. Yu, Q. Lin, S. Leung and Z. Fan, Nanomaterials and nanostructures for efficient light absorption and photovoltaics, Nano Energy 1, 57-72 (2012).

[33] L. Cao, J. S. White, J. S. Park, J. A. Schuller, B. M. Clemens and M. L. Brongersma, Engineering light absorption in semiconductor nanowire devices, Nat. Mater., 8, 643 (2009).

[34] H. Kallel, A. Chehaidar, A. Arbouet and V. Paillard, Enhanced absorption of solar light in Ge/Si core-sheath nanowires compared to Si/Ge core-sheath and $Si_{1-x}Ge_x$ nanowires: A theoretical study, J. Appl. Phys., 114, 224312 (2013).

[35] S. Sett, R. S. Bisht, A. Ghatak and A. K. Raychaudhuri, Surface oxide modification enables super-linear photoresponse in a single Germanium nanowire photodetector, Appl. Surf. Sc., 497, 143754 (2019)

[36] See Supplemental Material at [URL will be inserted by publisher] for Photocurrent versus wavelength plot in a single Ge nanowire.

[37] See Supplemental Material at [URL will be inserted by publisher] for absorption efficiency in a single Ge nanowire.

[38] See Supplemental Material at [URL will be inserted by publisher] for solution to heat diffusion equation in a cantilevered single Ge nanowire using finite element method.

[39] See Supplemental Material at [URL will be inserted by publisher] for determining the correction factor due to air convection.

[40] See Supplemental Material at [URL will be inserted by publisher] for Lorentzian fit to Raman peak.

[41] P. N. Martin, Z. Aksamija, E. Pop and U. Ravaioli, Reduced Thermal Conductivity in Nanoengineered Rough Ge and GaAs Nanowires, Nano Lett., 10, 1120-1124 (2010).

[42] J. M Ziman, Electrons and Phonons: The Theory of Transport Phenomena in Solids, (Clarendon Press, Oxford, UK).

[43] N. W. Ashcroft and N. D. Mermin, Solid State Physics, (Holt, Rinehart and Winston, New York 1976).

[44] S. Ghosh and A. K. Raychaudhuri, Link between depressions of melting temperature and Debye temperature in nanowires and its implication on Lindeman relation, J. Appl.Phys., 114, 224313 (2013).

[45] Y. Chen, D. Li, J.R. Lukes and A. Majumdar, Monte Carlo Simulation of Silicon Nanowire Thermal Conductivity, J. Heat Transfer, 127, 1129-1137 (2005).

[46] S. Voltz and G Chen, Molecular dynamics simulation of thermal conductivity of silicon nanowires, Appl. Phys. Lett., 75, 2056 (1999).

[47] Thermal Conductivity: Theory, properties and applications, Edited by T. M. Tritt (Kluwer Academic / Plenum Publishers, New York, 2004).

[48] J. P. Feser, J. S. Sadhu, B.P. Azeredo, K.H. Hsu, J. Ma, J. kim and M. Seong, Thermal conductivity of Silicon nanowire arrays with controlled roughness, J. Appl. Phys., 112, 114306 (2012).

[49] A. Kikuchi, A. Yao, I. Mori, I. Yamashita, T. Ono and S. Samukawa, Thermal conductivity of 10nm-diameter Silicon Nanowires Array Fabricated by Bio-Template and Neutral Beam Etching, IEEE 16th Int. Conf. Nanotech., 16, 505 (2016).

[50] Yunshan Zhao, Dan Liu, Jie Chen, Liyan Zhu, Alex Belianinov, Olga S. Ovchinnikova, Raymond R. Unocic, Matthew J. Burch, Songkil Kim, Hanfang Hao, Daniel S. Pickard, Baowen Li and John T. L. Thong, Engineering the thermal conductivity along an individual silicon nanowire by selective helium ion irradiation, Nat. Communications, 8, 15919 (2017).